\documentclass[prd,aps,twocolumn,a4paper,floatfix,nofootinbib]{revtex4}
   
\usepackage{graphicx,psfrag}
\usepackage{mathrsfs}
\usepackage{amsmath,amsfonts,amssymb}
\usepackage{multirow}
\usepackage{comment}
\usepackage[colorlinks=true]{hyperref}
\hypersetup{
  linkcolor=blue,
  citecolor=blue,
  urlcolor=blue
}
\usepackage{bbm}
\usepackage{placeins}
\usepackage[normalem]{ulem}
\usepackage{tensor} 
\usepackage{acronym}
\usepackage{xspace}
\usepackage{float}
\usepackage{booktabs,multirow,cellspace}

\usepackage{color}
\definecolor{cyan}{rgb}{0,0.9,0.9}
\definecolor{orange}{rgb}{0.9,0.5,0}
\definecolor{purple}{rgb}{0.8,0.4,0.8}
\definecolor{gray}{rgb}{0.8242,0.8242,0.8242}
\definecolor{pink}{rgb}{1.0, 0.0, 0.5}
\newacro{BBH}{binary-black-hole}
\newacro{BH}{black hole}
\newacro{BNS}{binary neutron star}
\newacro{EOB}{effective-one-body}
\newacro{EOS}{equation-of-state}
\newacro{GW}{gravitational-wave}
\newacro{NR}{numerical-relativity}
\newacro{NS}{neutron star}
\newacro{PN}{post-Newtonian}
\newacro{PSD}{power spectral density}

\newacro{SNR}{signal-to-noise ratio}

\definecolor{mygreen}{rgb}{0.1,0.8,0.1}

\begin{document}

\title{Investigating GW190425 with numerical-relativity simulations}

\author{Reetika \surname{Dudi$^1$}, 
Ananya \surname{Adhikari$^2$},  
Bernd \surname{Br\"ugmann$^2$}, 
Tim \surname{Dietrich$^{1,4}$},
Kota \surname{Hayashi$^{5}$},
Kyohei \surname{Kawaguchi$^6$},
Kenta \surname{Kiuchi$^{1,5}$},  
Koutarou \surname{Kyutoku$^{7,5,8}$}, 
Masaru \surname{Shibata$^{1,5}$}, 
Wolfgang \surname{Tichy$^2$}
}

\affiliation{$^1$ Max Planck Institute for Gravitational Physics (Albert Einstein Institute), Am Muhlenberg 1, Potsdam, Germany}
\affiliation{$^2$ Department of Physics, Florida Atlantic University, Boca Raton, FL  33431, USA} 
\affiliation{$^3$ Theoretical Physics Institute, University of Jena, 07743 Jena, Germany}
\affiliation{$^4$ Institut f\"{u}r Physik und Astronomie, Universit\"{a}t Potsdam, D-14476 Potsdam, Germany}
\affiliation{$^5$Center for Gravitational Physics, Yukawa Institute for Theoretical 
Physics, Kyoto University, Kyoto 606-8502, Japan}
\affiliation{$^6$Institute for Cosmic Ray Research, The University of Tokyo,
5-1-5 Kashiwanoha, Kashiwa, Chiba 277-8582, Japan}
\affiliation{$^7$Department of Physics, Kyoto University, Kyoto 606-8502, Japan}
\affiliation{$^8$Interdisciplinary Theoretical and Mathematical Science Program (iTHEMS), RIKEN, Wako, Saitama 351-0198, Japan}

\date{\today}

\begin{abstract} 
The third observing run of the LIGO-Virgo collaboration has resulted in about hundred gravitational-wave triggers including the binary neutron star merger GW190425. However, none of these events have been accompanied with an electromagnetic transient found during extensive follow-up searches. 
In this article, we perform new numerical-relativity simulations of binary neutron star and black hole - neutron star systems that have a chirp mass consistent with GW190425. Assuming that the GW190425's sky location was covered with sufficient accuracy during the electromagnetic follow-up searches, we investigate whether the non-detection of the kilonova is compatible with the source parameters estimated through the gravitational-wave analysis and how one can use this information to place constraints on the properties of the system.    
Our simulations suggest that GW190425 is incompatible with an unequal mass binary neutron star merger with a mass ratio $q<0.8$ when considering stiff or moderately stiff equations of state if the binary was face-on and covered by the observation. Our analysis shows that a detailed 
observational result for kilonovae will be useful to constrain 
the mass ratio of binary neutron stars in future events.
\end{abstract}


\maketitle

\section{Introduction}\label{Section:Introduction}

The increasing sensitivity of the Advanced LIGO~\cite{TheLIGOScientific:2014jea} and Advanced Virgo detectors~\cite{TheVirgo:2014hva} led to an increasing number of gravitational-wave (GW) detections.  
Among these, the first binary neutron star (BNS) detection during the third observing run (O3) was GW190425~\cite{Abbott:2020uma}, 
originally classified as S190425z. 
The total signal-to-noise ratio (SNR) of GW190425 was about 13.0 in GWTC-2 with an SNR of 12.9 in LIGO Livingston and only an SNR of 2.5 in Virgo which is below the detection threshold. 
Based on Refs.~\cite{Abbott:2020uma,TheLIGOScientific:2021gwc} the total mass of the system was about $3.4 M_\odot$ with component masses ranging between $1.12$ to $2.52 M_\odot$ (or $1.46–1.87 M_\odot $ if we restrict the dimensionless component spin magnitudes to be smaller than
$\chi \lesssim 0.05$). 
While these mass parameters are generally consistent with the possibility that both binary components are neutron stars, the source-frame chirp mass of $ 1.44^{+0.02}_{-0.02} M_\odot $ and the total mass of $3.4^{+0.3}_{-0.1} M_\odot$ of GW190425 are larger than those of any other known BNS systems (see, e.g., Ref.~\cite{Tauris:2017ytw}). 
Such massive systems will likely result in a prompt black hole (BH) formation once the two neutron stars merge~\cite{Shibata:2005ss,Shibata:2006nm,Kiuchi:2010ze,Hotokezaka:2011dh,Bauswein:2013jpa, Koppel:2019pys, Agathos:2019sah, Bauswein:2020xlt, Bernuzzi:2020tgt,Dietrich:2020eud}. 
Hence, the absence of an immediate detection of electromagnetic (EM) counterparts is not surprising, but could potentially still contain additional information about the source. 
Unfortunately GW190425, since it was a single detector event, was overall poorly localized, and the event was further away than GW170817~\cite{TheLIGOScientific:2017qsa}, which created an additional challenge for EM campaigns.
Nevertheless, as the first alert during O3 with a high probability of having a counterpart, there was an intense campaign within the first few days after the initial alert and about $120$ GCNs have been created. 
Among them, the `Global Relay of Observatories Watching Transients Happen' network observed thousands of square degrees, which amount to 46\% and 21\% of the 90\% probability region derived by BAYESTER and LALInference, respectively. It has also been reported that GRB 190425 coincident with GW190425 had come from the northern hemisphere~\cite{Coughlin:2019grw}, which, however, might be a coincidence.
Similarly, GOTO also covered up to 30\% of the initial sky map down to a magnitude of about 20.5~\cite{steeghs:2019got}. A summary of most of the observations can be found in Table 2 of Ref.~\cite{Coughlin:2019avz}.

\begin{table*}[t]
\caption{BNS and BHNS models employed in this paper.
The first column gives the model name.
The next 5 columns provide the physical properties of the individual stars:  EOS, the gravitational masses of the individual stars $M^{A,B}$, and the baryonic masses of
the individual stars $M _b ^{A,B}$. The last 9 columns give mass ratio $q$, 
 the initial GW
frequency $M\omega^0 _{22}$, the Arnowitt-Deser-Misner (ADM) mass $M_\text{ADM}$, the  angular
momentum $J$, the compactness of the stars $C^{A,B}$ and their dimensionless tidal deformability $\Lambda_2^{A,B}$. The BNS models are evolved with the grid resolutions of Table~\ref{tab:grid}.
}
\label{tab:config}
\begin{tabular}{c|cccccccccccccccc}
\toprule
Name & EOS &$M^{A}$ & $M^{B}$ & $M^{A}_b$ & $M^{B}_b$  & $q$ & $M\omega ^0 _{22}$ &  $M$ &$M_\text{ADM}$ & $J$ & $C^A$ & $C^B$ & $\Lambda_2^A$ & $\Lambda_2^B$\\
\hline
APR4$_{0.7}$ & APR4  & 1.982 & 1.388 & 2.355 & 1.550 & 0.7 &   0.0599 & 3.371 & 3.336 &  10.419 & 0.2694 & 0.1850 & 17.3 & 271 \\
APR4$_{0.8}$ & APR4 & 1.852 & 1.481 & 2.166  & 1.668 & 0.8 &  0.0590  & 3.333 & 3.297 & 10.419 & 0.2493 & 0.1973 & 33.6 & 180 \\ 
APR4$_{0.9}$ & APR4 & 1.744 & 1.569 & 2.016  & 1.783  & 0.9 &  0.0586 & 3.312 & 3.279  & 10.414 & 0.2331 & 0.2091 & 55.4 & 121\\ 
APR4$_{1}$ & APR4 & 1.654  &  1.654 & 1.894 & 1.894 & 1.0 & 0.0584 & 3.308 & 3.273 & 10.405 & 0.2204 & 0.2204 & 83.2 & 83.2 \\ 
 \hline \hline
 
DD2$_{0.7}$ & DD2  & 1.982 & 1.391 & 2.271 & 1.519 & 0.7 &   0.0599 & 3.373 & 3.338 &  10.416 &  0.2220 &  0.1152 &  74.2 & 735 \\ 
DD2$_{0.8}$ & DD2 & 1.862 & 1.467 & 2.110  & 1.611 & 0.8 &  0.0589  & 3.329 & 3.294 & 10.37 & 0.2077 &  0.1634 &  118 &  542\\ 
DD2$_{0.9}$ & DD2 & 1.744 & 1.565 & 1.958  & 1.732  & 0.9 &  0.0584 & 3.309 & 3.274  & 10.37 &  0.1941&  0.1742 &  186 & 368 \\ 
DD2$_{1}$ & DD2 & 1.655  &  1.655 & 1.845 & 1.845 & 1.0 & 0.0585 & 3.311 & 3.275 & 10.42 &  0.1841&  0.1841 &  260 &  260 \\ 
 \hline \hline

MPA1$_{0.7}$ & MPA1  & 1.983 & 1.388 & 2.310  & 1.537 & 0.7 & 0.060   & 3.372  & 3.337  & 10.474 & 0.2389 & 0.1692 & 50.2 & 519 \\ 
MPA1$_{0.8}$ & MPA1 & 1.852 & 1.481 & 2.131  & 1.651 & 0.8 & 0.0591   & 3.333 & 3.298 & 10.447 & 0.2229 & 0.1797 & 83.6 & 357 \\ 
MPA1$_{0.9}$ & MPA1 & 1.744 & 1.569 & 1.987  &  1.762 & 0.9 &  0.0586 & 3.313 & 3.279  & 10.444 & 0.2100 & 0.1899 & 127 & 251 \\ 
MPA1$_{1}$ & MPA1 & 1.655  &  1.655 & 1.872 & 1.872 & 1.0 & 0.0586 & 3.309 & 3.310 & 10.459 & 0.1998 & 0.1998 & 179 & 179 \\ 
\hline \hline 
BHNS$_{0.7}$ & APR4  & 1.980 & 1.390 & - & 1.552 & 0.7 & 0.0560 & 3.370 & 3.337 &  10.582  & - & 0.1851 & 0 & 268 \\
 \hline \hline
 
\end{tabular}
\end{table*}

There have already been several studies related to GW190425, which explore the possibility that this event might arise from a black hole-neutron star (BHNS), not from a BNS merger  (e.g., Refs.~\cite{Han:2020qmn,Kyutoku:2020xka}), its implication for the neutron star equation of state (e.g., Refs.~\cite{Agathos:2019sah,Landry:2020vaw,Dietrich:2020efo,Huth:2021bsp}), possible implications from the absence of an EM counterpart (e.g., Ref.~\cite{Coughlin:2019avz}), and implications with respect to the population of BNSs~\cite{Abbott:2020uma}. In this paper, we perform numerical-relativity simulations to invesigate (i) how consistent the absence of an EM counterpart with the estimated GW source parameters is and (ii) whether the non-detection of an EM counterpart can be used to constrain the binary parameters assuming that the event was in the observed area. 
For this purpose, we simulate twelve different BNS setups with fixed chirp mass $(\mathcal{M} = 1.44 M_\odot)$ as estimated for GW190425. Our configuration spans the mass ratios\footnote{ $q = M^B/M^A \leq 1$, with $M^{A,B}$ being the individual gravitational masses of stars in isolation.} of $q=0.7$, 0.8, 0.9, and 1 and, for each mass ratio, we employ three equations of state (EOSs), namely, APR4, DD2, and MPA1. We neglect intrinsic NS spin focusing only on non-spinning configurations.
In addition, we simulate one BHNS system with the mass ratio $q=0.7$ and the APR4 EOS for comparison about the merger properties and gravitational waveform.

Throughout this work, we basically use geometric units, setting $c = G = M_\odot = 1$. 
However, we sometimes include $M_\odot$ explicitly or quote values in cgs units 
for the case that the units should be clarified.

\section{Methods and Configurations}

In this article, we simulate twelve different BNS configurations and one BHNS configuration. Table~\ref{tab:config} provides a detailed list of the initial properties for all the models. 
All the BNS systems are simulated using the \texttt{BAM} code~\cite{Brugmann:2008zz, Thierfelder:2011yi, Dietrich:2018phi}, where we employ the Z4c scheme~\cite{Bernuzzi:2009ex,Hilditch:2012fp} along with the 1+log and gamma-driver conditions for the evolution of the lapse and shift \cite{Bona:1994dr,Alcubierre:2002kk,vanMeter:2006vi}. All the models are studied with three different grid resolutions consisting of 7 refinement levels which represent a hierarchy of cell centered nested Cartesian grids, three out of which are dynamically moving; cf.~Tab.~\ref{tab:grid} for further details about the BAM grid setup. 
All the outputs, such as the GW signal, ejecta and disk mass properties, and 
remnant BH quantities, are extracted using standard techniques described in Refs.~\cite{dbt_mods_00029293, Dietrich:2016hky, Chaurasia:2018zhg}. 

BNS initial data are constructed with the pseudo-spectral \texttt{SGRID} code~\cite{Tichy:2009yr, Tichy:2011gw, Tichy:2012rpi, Tichy:2016vmv, Dietrich:2015iva, Tichy:2019ouu}, where we use $n_x = 36$, $n_y=36$, and $n_z=36$ points for the spectral grids. We perform an eccentricity reduction procedure as described in Ref.~\cite{Tichy:2019ouu} to obtain residual eccentricity of about $\leq 10^{-3}$ for all our initial data.

The BHNS simulations are performed with the \texttt{SACRA} code. 
\texttt{SACRA} solves the Einstein equation in a moving puncture version of the BSSN formulation, but incorporates a Z4c constraint-propagation prescription.
We refer the reader to Refs.~\cite{Kyutoku:2015gda, Kawaguchi:2015bwa, Kiuchi:2017pte} for a detailed discussion of the employed formulations, initial condition, and diagnostics for the latest version of \texttt{SACRA}. 

The EOSs used in this article are as follows: (i) APR4, derived by a variational-method with the AV18 2-body potential, the UIX 3-body potential, and relativistic boost corrections~\cite{Akmal:1998cf}. It supports the maximum mass of non-rotating TOV star with about $M_{\rm tov} \approx$ 2.21 $M_\odot$ and the radius of $1.35M_\odot$ NS (referred to as $R_{1.35}$) is $\approx 11.4$\,km; 
(ii) a piecewise polytropic representation of the tabulated DD2 EOS~\cite{Typel:2009sy}, which we fit for our simulations with a five piece piecewise polytrope. It supports $M_{\rm tov} \approx$ 2.42 $M_\odot$, and $R_{1.35}\approx 13.2$\,km; and 
(iii) MPA1 which is a piecewise-polytropic EOS that supports $M_{\rm tov} \approx$ 2.47 $M_\odot$ 
with $R_{1.35}\approx 12.5$\,km~\cite{Read:2008iy}. For all the EOSs, 
the value of $M_{\rm tov}$ is fairly high $\agt 2.2M_\odot$, but irrespective of 
the EOSs and mass ratios, the BNS models employed result in a black hole 
in a short time scale after the onset of merger 
because of the high total mass of the system 
$\approx 3.3$--$3.4M_\odot > 1.3M_{\rm tov}$ (see the next section).

\begin{table}
\caption{Grid configurations for the BAM simulations. The columns refer to: the resolution name,
the total number of refinment levels $L$, the number of moving box levels $L_{\rm mv}$,
the number of points in the non-moving boxes $n$, the number of points
in the moving boxes $n_{\rm mv}$, the grid spacing in the finest level
$h_6$ covering the NS diameter, the grid spacing in the coarsest level
$h_0$, and the outer boundary position along each axis $R_0$. The grid spacing and the
outer boundary position are given in units of $GM_{\odot}/c^2 \approx 
1.477\,\mathrm{km}$.}
\label{tab:grid}
\centering
\begin{tabular}{cccccccc}
\toprule
Name & $L$ & $L_{\rm mv}$ & $n$ & $n_{\rm mv}$ & $h_6$ & $h_0$ & $R_0$
\\ \hline
R1 & 7 & 3 & 384 & 192 & 0.070 & 4.467 & 859.83\\
R2 & 7 & 3 & 288 & 144 & 0.093 & 5.956 & 860.50 \\
R3 & 7 & 3 & 192 & 96  & 0.140 & 8.933 & 862.06\\ \hline \hline
\end{tabular}
\end{table}

\section{BNS Results}

\subsection{Gravitational Waves}
\begin{figure}[t]
\includegraphics[width=0.5\textwidth]{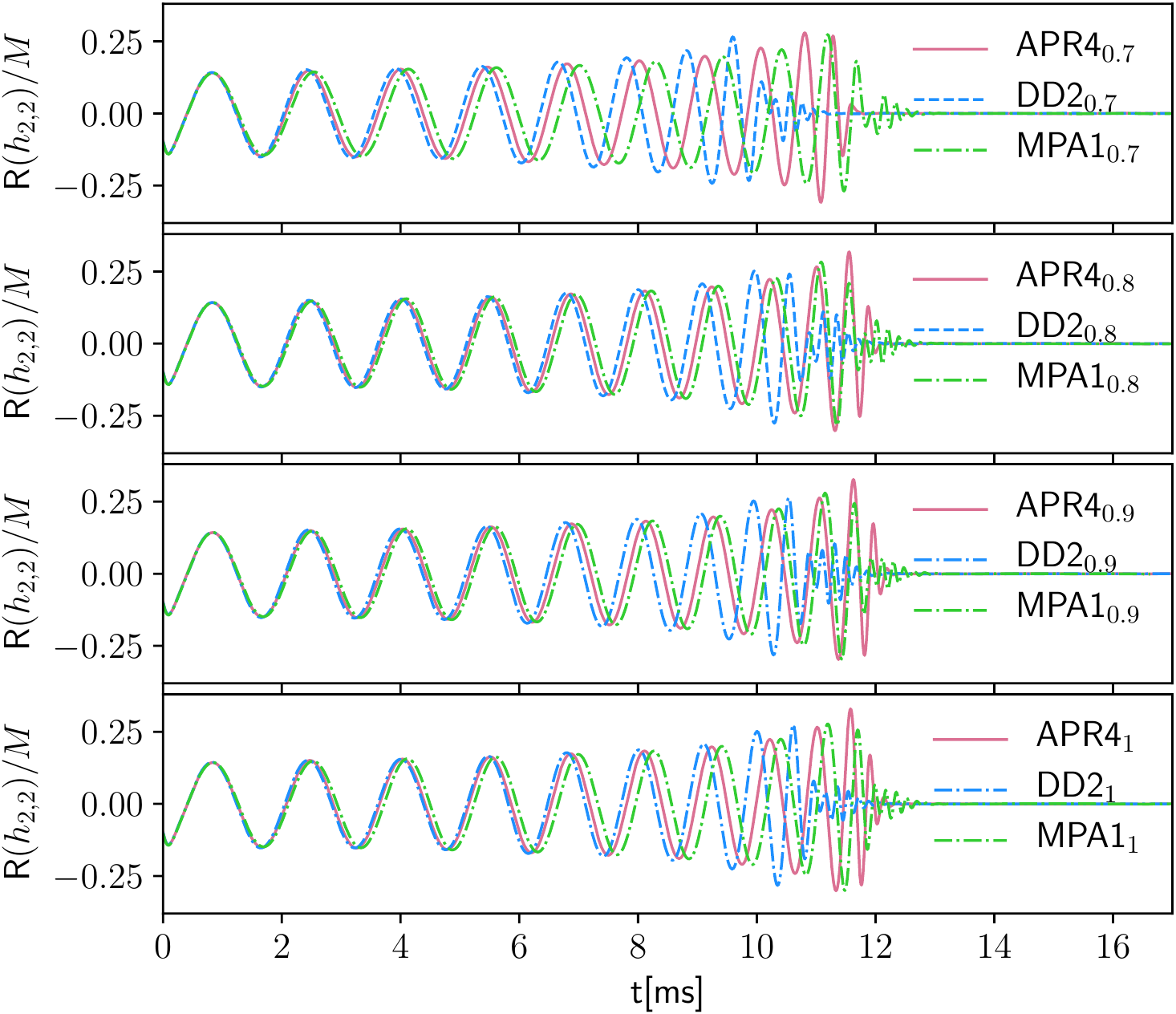}
\caption{The $(2,2)$-mode of GWs for our BNS simulations is displayed. 
The results with the APR4, DD2, and MPA1 EOSs are shown in pink, blue, and green, respectively. The individual panels show from top to bottom 
the results with different mass-ratio, namely, $q =0.7, 0.8, 0.9$, and $q=1$, respectively. }
\label{fig:GW}
\end{figure}

All the BNS models with the high total mass employed in our simulation (Tab.~\ref{tab:config}) lead to quick BH formation after the onset of merger. 
This is visible in Fig.~\ref{fig:GW}, which shows a quick damping of the 
gravitational-wave amplitude of the dominant $(2,2)$-mode for 
all the BNS models in the merger and post-merger phases. 
This finding is consistent with recently published results; e.g., Refs.~\cite{Bauswein:2020xlt, Agathos:2019sah}. 
Nevertheless, slight differences between the individual setups are appreciable. 
Most notably, setups evolved with the MPA1 and DD2 EOS show small oscillations after the peak in the GW amplitude, 
because the merged object survives for a few milliseconds before the collapse to a BH. 
However, such small short-term oscillations have a high frequency of 2--3\,kHz, and thus, 
they 
would not be detectable with the existing 2 and 2.5 generation of GW detectors because of their poor sensitivity in the 
high-frequency band with $\agt 2$\,kHz.

We note that, although the dephasing in the inspiral phase shown in Fig.~\ref{fig:GW} should result primarily from different binary tidal deformability \cite{Flanagan:2007ix,Hinderer:2009ca},
\begin{equation}
\tilde{\Lambda}_2 = \frac{16}{13} 
\frac{(1+12 q) \Lambda_2^A + (q+12)q^4 \Lambda_2^B}{(1+q)^5},
\end{equation}
we also see deviations from the expected dependence. Specifically, the values of $\tilde{\Lambda}_2$ is $85.5$, $178.8$, $259.7$ for APR4$_{0.7}$, MPA1$_{0.7}$, and DD2$_{0.7}$, respectively, and binaries with large values of $\tilde{\Lambda}$ are expected to evolve faster. However, the inspiral of MPA1 setups is systematically slower than the expected trend. This stems from the excess angular momentum in the initial data for MPA1 setups caused by the limited computational accuracy. Because the binary in MPA1 setups needs to emit larger amount of angular momentum than other setups, the inspiral phase happens to last longer.

\begin{table}
\caption{Properties of the merger remnant.
The columns represent:
(i) the name of the configuration;
(ii) the final mass of the BH, $M_\text{BH}$, 
the dimensionless spin of the final BH, $\chi_\text{BH}$,
and the mass of the disk surrounding the BH,
$M_\text{disk}$;
(iii) the ejecta mass $M_{\rm ej}$. 
 The different physical quantities are computed for all the resolutions (R1, R2, R3)
 (from top to bottom), respectively.
 }
\label{tab:ejecta}
\centering

\begin{tabular}{c|ccc|c}
\toprule
Name  & ~$M_\text{BH}$~ & ~$\chi_\text{BH}$~ & ~$M_\text{disk}$~ & $M_\text{ej}[M_{\odot}$] \\
    &   \begin{footnotesize} $[M_{\odot}]$  \end{footnotesize}& & \begin{footnotesize} $[M_{\odot}]$  \end{footnotesize}&                   \\
\hline
\multirow{3}{*}{APR4$_{0.7}$} & 3.284  & 0.738 & $4.8 \times 10^{-3}$ & $5.2 \times 10^{-4}$   \\
                              & 3.284 & 0.729 & $3.0 \times 10^{-3}$& $3.0\times 10^{-4}$    \\
                              &  3.276 & 0.729 & $3.4\times 10^{-3}$ & $9.0\times 10^{-4}$  \\

\hline

\multirow{3}{*}{DD2$_{0.7}$}  & 3.246  & 0.732 & $5.6 \times 10^{-2}$& $2.4 \times 10^{-3}$  \\
                              & 3.240  & 0.724 & $5.4\times 10^{-2}$ & $4.5 \times 10^{-3}$  \\
                              & 3.207 & 0.692 & $0.8\times 10^{-2}$ & $8 \times 10^{-3}$   \\

\hline
\multirow{3}{*}{MPA1$_{0.7}$} & 3.259  & 0.738 & $3.8\times 10^{-2}$ & $1.1 \times 10^{-2}$  \\
                              & 3.259  & 0.737 & $3.8\times 10^{-2}$  & $1.1 \times 10^{-2}$   \\
                              & 3.252  & 0.728 & $3.8 \times 10^{-2}$ & $1.2\times 10^{-2}$ \\

\hline
\hline

\multirow{3}{*}{APR4$_{0.8}$}  &    3.245 & 0.730 & $3.4 \times 10^{-4}$ &  $1.4\times 10^{-4}$  \\
                               &  3.244 & 0.719 & $2.6\times 10^{-4}$ & $9.3\times 10^{-5}$  \\
                               & 3.240 & 0.720  & $1.6 \times 10^{-3}$& $2.2\times 10^{-4}$ \\
\hline

\multirow{3}{*}{DD2$_{0.8}$}  &    3.209 & 0.728 & $3.3\times 10^{-2}$ &  $5.5 \times 10^{-3}$  \\
                               &  3.209 & 0.727 & $3.0\times 10^{-3}$ & $8.8\times 10^{-3}$  \\
                               & 3.180 & 0.695  & $2.3\times 10^{-3}$ & $1.3\times 10^{-2}$ \\
\hline
\multirow{3}{*}{MPA1$_{0.8}$} & 3.247  & 0.755 &  $8.6\times 10^{-3}$ &  $3.5\times 10^{-3}$ \\
                              & 3.246  & 0.752 & $8.7 \times 10^{-3}$ & $2.8 \times 10^{-3}$   \\
                              & 3.244  & 0.748 & $5.2\times 10^{-3}$ &  $4.8 \times 10^{-3}$ \\

\hline
\hline 

\multirow{3}{*}{APR4$_{0.9}$}  & 3.222 & 0.723 &  $<10^{-6}$ & $<10^{-6}$  \\
                               & 3.220 & 0.724 & $3.8 \times 10^{-5}$   & $<10^{-6}$  \\
                               &  3.219  &  0.719  & $4.6 \times 10^{-5}$ & $<10^{-6}$  \\

\hline
\multirow{3}{*}{DD2$_{0.9}$}  & 3.224 & 0.754 & $3.4 \times 10^{-3}$ & $1.8 \times 10^{-3}$\\
                               & 3.252 &   0.747 & $6.3 \times 10^{-4}$ & $3.8 \times 10^{-3}$  \\
                               & 3.192  &  0.721 & $<10^{-6}$  & $3.1 \times 10^{-3}$  \\

\hline
\multirow{3}{*}{MPA1$_{0.9}$} & 3.234  & 0.756 & $2.2\times 10^{-3}$ &  $1.6\times 10^{-3}$  \\
                              & 3.233  & 0.755 & $1.6 \times 10^{-3}$& $0.9 \times 10^{-3}$   \\
                              & 3.230  & 0.749 & $6.2\times 10^{-4}$ & $0.4 \times 10^{-3}$ \\

\hline
\hline

\multirow{3}{*}{APR4$_{1}$}  &   3.215  & 0.721 &  $<10^{-6}$ & $0.2 \times 10^{-5}$\\
                              &  3.215  & 0.722 &  $<10^{-6}$ & $0.2\times 10^{-5}$ \\
                               &   3.214  & 0.719 &  $<10^{-6}$ & $2.5 \times 10^{-5}$  \\

\hline 

\multirow{3}{*}{DD2$_{1}$}  &   3.233  & 0.769 & $<10^{-6}$ & $8.1 \times 10^{-4}$ \\
                              &   3.223 & 0.758  & $<10^{-6}$ & $8.5 \times 10^{-4}$   \\
                               &   3.202  & 0.734 & $<10^{-6}$ &  $<10^{-6}$  \\

\hline 
\multirow{3}{*}{MPA1$_{1}$} & 3.231  & 0.757 &  $<10^{-6}$ &   $0.5 \times 10^{-3}$ \\
                              & 3.232  & 0.757  & $<10^{-6}$ &  $0.4 \times 10^{-3}$  \\
                              & 3.226  & 0.748 & $<10^{-6}$ & $0.7 \times 10^{-4}$  \\

\hline
\hline 

\end{tabular}

\end{table}

\subsection{Dynamical Ejecta Properties}
\begin{figure*}
\centering
\includegraphics[width=0.8\textwidth]{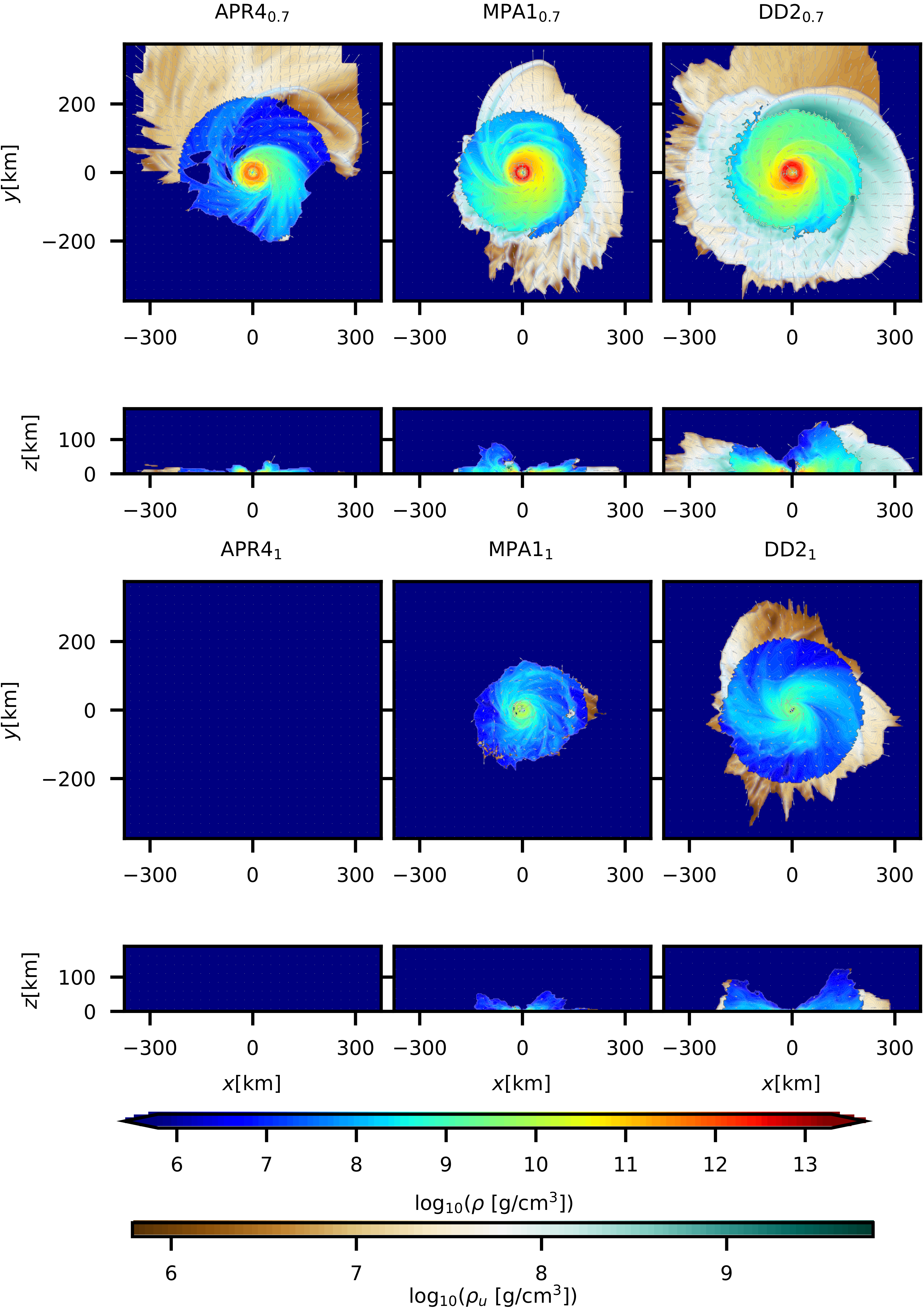}
\caption{Rest-mass density profile and velocity field inside the orbital plane for all simulations with $q=1$ and $0.7$. The snapshots represent the moments at the merger. The rest-mass density ($\rho$ in units of ${\rm g/cm}^3$) is shown on a logarithmic scale from blue to red. The rest-mass density of unbound material ($\rho_u$) is colored from brown to dark green.}
\label{fig:ejecta}
\end{figure*}

We compute the dynamical ejecta mass for all our simulation models, and provide all information about it in Tab.~\ref{tab:ejecta} (see also Fig.~\ref{fig:ejecta}).
We find that -- in agreement with previous studies, e.g.,~Refs.~\cite{Hotokezaka:2012ze,Bauswein:2013yna,Dietrich:2015iva} -- unequal-mass systems eject more material during the merger phase, 
as compared to equal-mass mergers. This is caused by the large tidal distortion of the light component right before the contact. At this point, the tidal interaction effectively ejects an outer part of the light component in an unequal-mass binary. Furthermore, the dynamical ejecta mass is higher for simulations with stiffer EOSs (MPA1, DD2) as compared to soft EOS (APR4). 
Our findings indicate that the main ejection mechanism in these high-mass mergers is tidal dynamical ejecta as discussed in, e.g.,~Refs.~\cite{Kiuchi:2019lls,Bernuzzi:2020txg,Bernuzzi:2020tgi}. 
The shock heating effect at the merger plays a role in the mass ejection 
for the nearly equal-mass case~\cite{Hotokezaka:2012ze}. However, this 
effect is important only for soft EOSs, and thus, in the present 
context, the effect plays only a minor role, because in the model with the soft EOS (i.e., 
APR4), nearly all the NS matter collapses promptly into the black hole at the merger (see Fig.~\ref{fig:ejecta}).

\subsection{Remnant Properties}
\begin{figure}[t]
\includegraphics[width=0.5\textwidth]{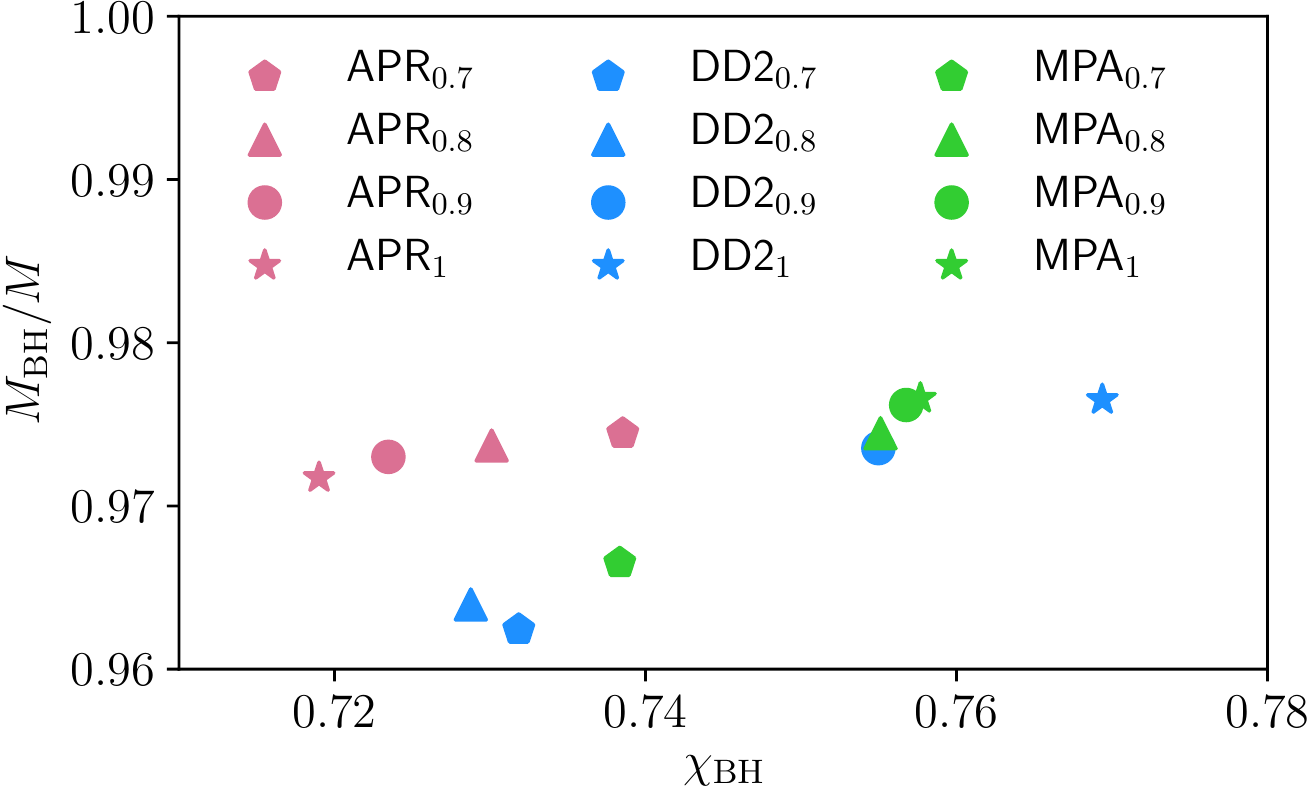}
\caption{Mass and spin of the remnant BH. The pink, blue, and green show the results with APR4, DD2, and MPA1 EOSs, respectively. Different symbols denote models with different mass ratios, i.e., $q=0.7$ by the pentagon, $q=0.8$ by the triangle, 
$q=0.9$ by the circle, and $q=1$ by the star.}
\label{fig:chibh}
\end{figure}

Table~\ref{tab:ejecta} and Fig.~\ref{fig:chibh}  
show the final mass and spin of the remnant BH. 
In general, differences of the final BH mass and spin are caused by (i) the different amount of emitted GWs, i.e., more compact stars tend to merge later and therefore can release more energy and angular momentum via GWs, and (ii) a large amount of mass and angular momentum can be extracted from the system via the ejection of material or could be stored in the remnant disk surrounding the final BH.

For the models with stiffer EOSs and lower mass ratio like MPA1$_{0.7}$ and DD2$_{0.7}$, we find that the dynamical ejecta mass and the disk mass are noticeably higher than for 
equal- or nearly equal-mass systems (i.e., $q=1$ or 0.9) 
so that the total mass and angular momentum of the final BH are reduced.  
This explains that for MPA1 and DD2, asymmetric systems produce lighter BHs with smaller spin. 
The opposite is true for the APR4 setup, where the equal-mass system 
produces the smallest BH mass and spin. 
Our interpretation for this result is that although the masses of the dynamical 
ejecta and remnant disk are smaller for the more symmetric systems ($q\rightarrow 1$), 
more energy and angular momentum are dissipated by the GW emission and 
this effect plays a more important role for determining the property of 
the remnant BH.


\subsection{Lightcurve Computation}
We follow the methods outlined in Refs.~\cite{Tanaka:2013ana,Tanaka:2017qxj,Tanaka:2017lxb,Kawaguchi:2019lv} to compute the kilonova lightcurves: 
We do not simulate the evolution of the electron fraction and of the internal composition of the fluid, and rely on simplified models for estimating the lightcurves. In particular we use the ejecta profile of Ref.~\cite{Kawaguchi:2019lv}, i.e., we assume that any kind of post-merger ejecta is slower than the dynamical ejecta.

The kilonova lightcurves are calculated for the BNS models using the dynamical ejecta mass and the disk mass summarized in Tab.~\ref{tab:ejecta}. In this calculation, 
we assume that 20\% of the reported disk mass gets ejected following the latest 
results (see, e.g., Ref.~\cite{Fujibayashi2020a}). We have verified that our conclusions about the detectability of kilonovae does not change for a reasonable range of the ejection efficiency and confirmed that the variation of the ejection efficiency within 10\%--30\% modifies the magnitude of kilonovae only by $\sim 0.5\,\mathrm{mag}$; see also,  e.g.,~Refs.~\cite{Fernandez:2014cna,Siegel:2017jug,Fernandez:2018kax,Coughlin:2019zqi} for studies about the efficiency of ejecting part of the remnant disk. 
Because the electron fraction, and hence, the lanthanide fraction of the post-merger ejecta depend on the time scale of mass ejection~\cite{Fujibayashi:2020dvr,Fujibayashi2020a}, we calculate two types of models for each BNS model; one assuming the lanthanide-poor post-merger ejecta ($Y_e = 0.3$--$0.4$) and the other assuming the mildly lanthanide-rich post-merger ejecta ($Y_e = 0.2$--$0.4$).



\begin{figure*}[t]
\includegraphics[width=1.0\textwidth]{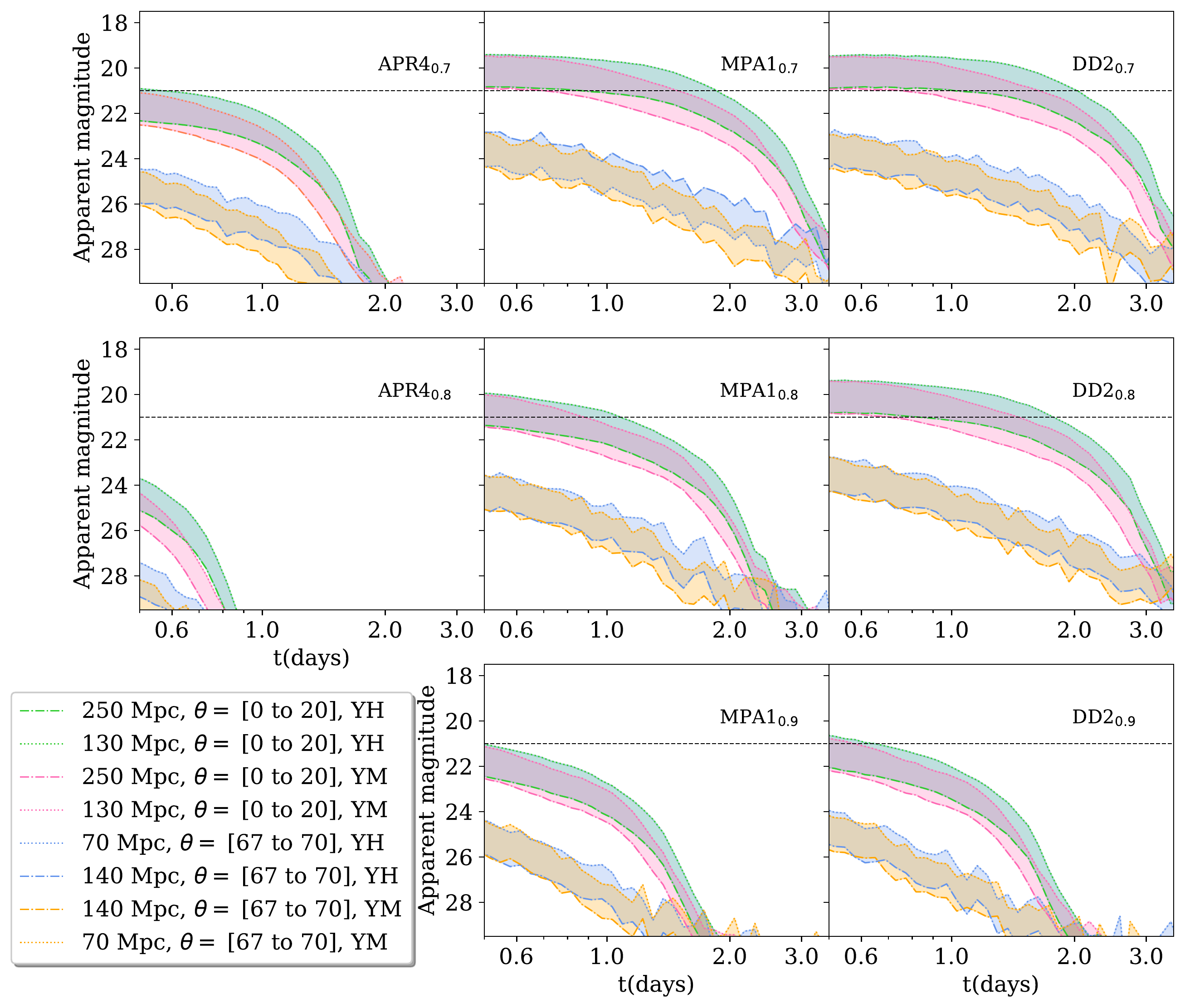}
\caption{Apparent AB magnitude for the r-band lightcurve of the kilonova models for GW190425-like events. EOSs are varied from left to right, i.e, simulations with APR4 (left panel), MPA1 (middle panel), and DD2 (right panel), and mass-ratio is varied from top to bottom, i.e., simulations with $q=0.7$ (top panel), $q=0.8$ (middle panel), and $q=0.9$ (bottom panel). Pink and green bands represent variations in the distance (from 130 Mpc to 250 Mpc) and orientation ($0^\circ \leq \theta \leq 20^\circ $), while orange and blue bands represent variations in the distance from 70 Mpc to 140 Mpc and $67^\circ \leq \theta \leq 70^\circ$. We made two assumptions for the electron fraction; lanthanide-poor post-merger ejecta (YH) and lanthanide-rich post-merger ejecta (YM).}
\label{fig:lightcv}
\end{figure*}

Figure~\ref{fig:lightcv} shows the apparent magnitudes for the $r$-band lightcurve of the GW190425-like kilonova events varying the EOSs: APR4 (left panel), DD2 (middle panel), and MPA1 (right panel), and the mass ratio: $q=0.7$ (top panel), $q=0.8$ (middle panel), and $q=0.9$ (bottom panel)\footnote{In the model 
APR4$_{0.9}$, material is not ejected appreciably enough to provide realistic lightcurve estimates.}. 
The pink bands are produced by varying the distance between 250 Mpc to 130 Mpc,  ($0^{\circ} < \theta <20^{\circ} $), and assuming the lanthanide-rich post-merger ejecta (YM). 
On the other hand, the green bands are produced by varying the distance and $ \theta$ similarly, but assuming he lanthanide-poor post-merger ejecta (YH). 
The orange bands are produced by varying the distance between 140 Mpc to 70 Mpc,  ($67^{\circ} < \theta <70^{\circ} $) and assuming the lanthanide-rich post-merger ejecta (YM), and the blue bands by assuming the lanthanide-poor post-merger ejecta (YH). 
We have picked up these different inclination angles and distances based on Fig.~11 of Ref.~\cite{Abbott:2020uma}, which shows the large degeneracy in the estimation for the distance and inclination angle of GW190425. The black dashed lines indicate 21 mag, as the upper-limit at $\approx 1$ day after the trigger for the observation during the follow-up campaign~\cite{Coughlin:2019grw}. 

\begin{figure}[t]
\includegraphics[width=0.45\textwidth]{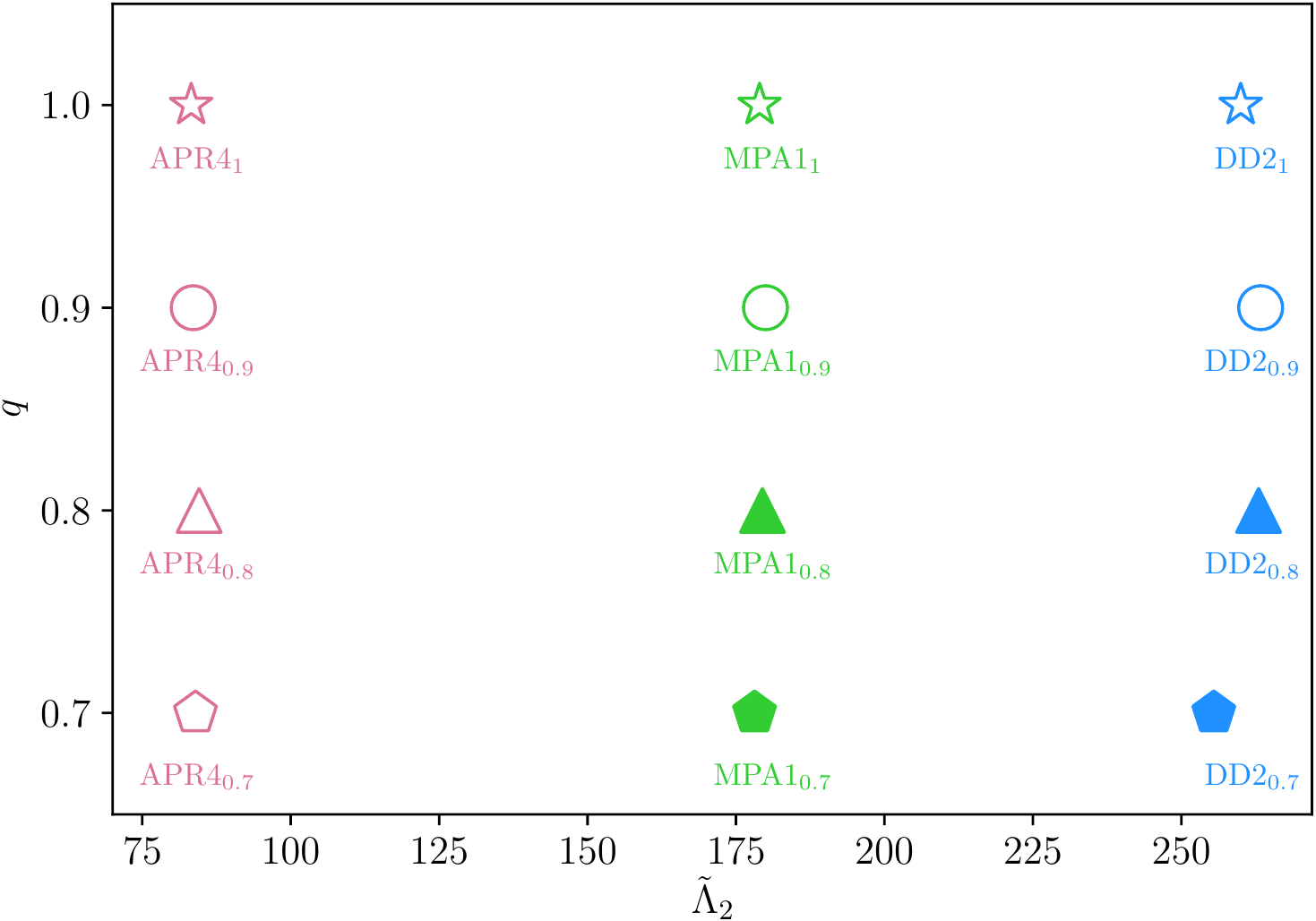}
\caption{Mass-ratio ($q$) vs. tidal deformability ($\tilde{\Lambda}_2$) for all our models. 
The filled symbols denote the models by which a possible kilonova brighter than 21 mag could be produced and the open symbols by which non-detectable signal was expected. }
\label{fig:q-lam}
\end{figure}


Our analysis allows us to rule out the models with lightcurves that reach magnitudes brighter than 21\,mag, given the observational constraints of Ref.~\cite{Coughlin:2019grw} under the assumption that the location of GW190425 was covered by the follow-up. It is found that systems like DD2$_{0.7}$, DD2$_{0.8}$, MPA1$_{0.7}$, and MPA$_{0.8}$ for $0^{\circ} < \theta <20^{\circ}$ and $D \in [130,250]\rm Mpc$ are disfavored. 
By contrast, it is difficult to rule out any EOS model if the mass ratio is 
close to unity with which the masses of the dynamical ejecta and remnant disk are small. For $67^{\circ} < \theta <70^{\circ}$ and $D \in [70,140]\rm Mpc$, all the cases are consistent with the non-detection of the kilonova, and hence, no information about the EOS or the mass ratio could be obtained. This is due to the presence of the neutron-rich dynamical ejecta, as we explain in more detail in the next section. 

We summarize this finding in Fig.~\ref{fig:q-lam}, in which we report all our BNS 
simulation models in the plane of tidal deformability and mass ratio. It is evident that the systems with larger tidal deformability and more asymmetric mass could produce kilonova lightcurves that might have been detectable; in other words, the non-detection of kilonova signals provides additional information beyond the GW signal.

\section{BNS vs.\ BHNS comparison}

To recall, GW observations with current sensitivities are unlikely to differentiate between BHNS binaries and BNSs, but such a differentiation will become possible with future sensitivity upgrades, e.g.,~Refs.~\cite{Chen:2020fzm,Fasano:2020eum}. Even after detecting an EM counterpart for GW170817, it has not been strictly confirmed to be a BNS merger~\cite{TheLIGOScientific:2017qsa,Hinderer:2018pei,Coughlin:2019kqf}. With respect to GW190425 the situation is even more challenging, because no associated EM counterpart was found, and due to the high total mass of the system, tidal effects were not appreciable; cf.~Refs.~\cite{Han:2020qmn,Kyutoku:2020xka} for a discussion about a possible BHNS origin of GW190425. In this section, we indicate that the future kilonova observations could 
help to distinguish between BNS and BHNS mergers even for a high-mass system like GW190425 although it is quite difficult to do so only from the GW observation. We here 
follow up the discussions in Ref.~\cite{Kyutoku:2020xka} by computing lightcurves for BNS and BHNS mergers based on numerical-relativity simulations tailored to the parameters of GW190425.

\subsection{Comparison of numerical-relativity simulations}

\begin{figure}[t]
\includegraphics[width=0.45\textwidth]{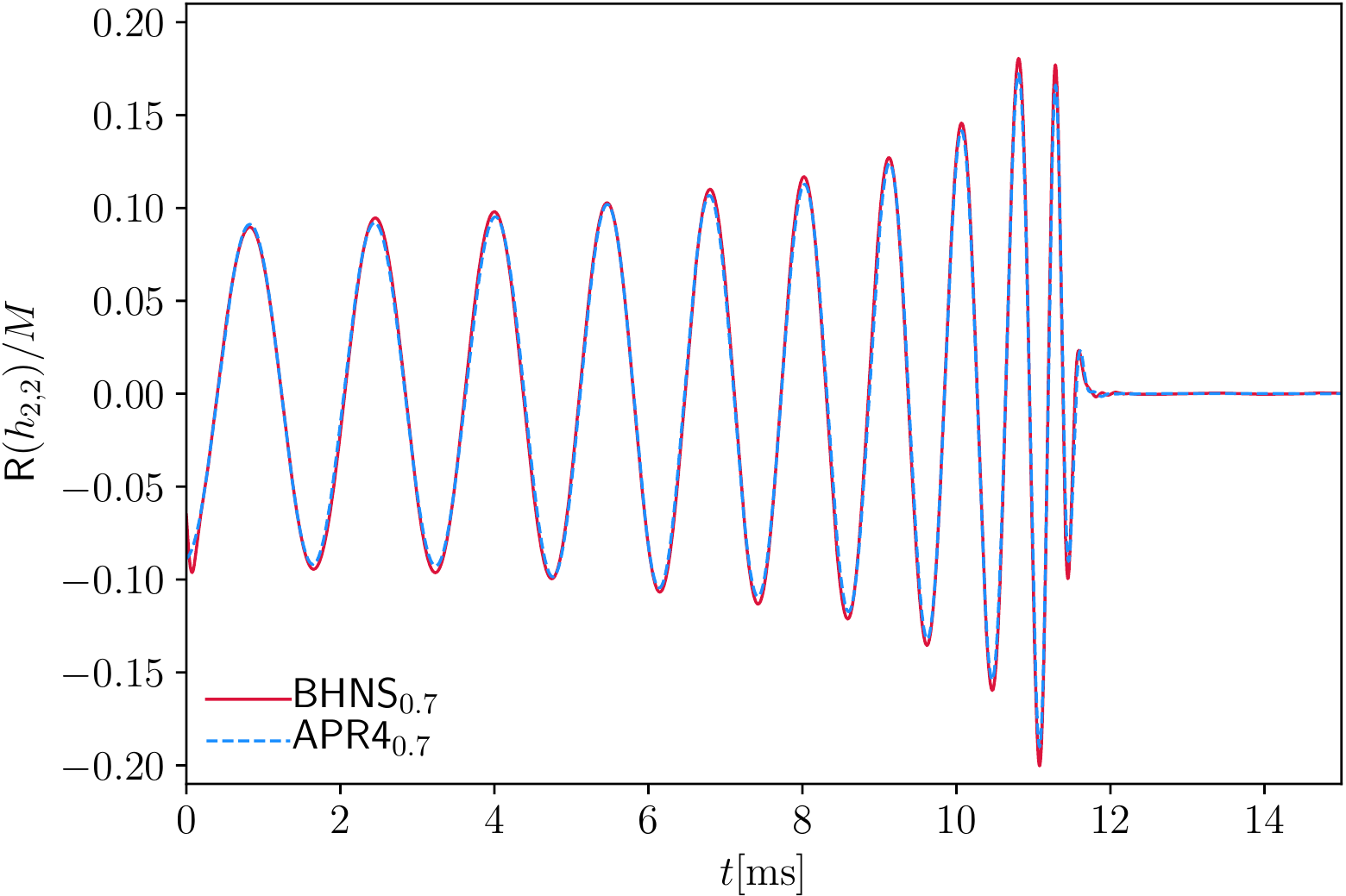}
\caption{(2,2)-mode of the GW strain for the BHNS model (BHNS$_{0.7}$: red solid curve) and for a BNS model (APR4$_{0.7}$: blue dashed curve). Two waveforms are aligned in time and phase to maximize the match. We emphasize, as written in the main text, that this good agreement results from the very small difference in the tidal deformability $\tilde{\Lambda}_2=71.2$ for BHNS$_{0.7}$ and $\tilde{\Lambda}_2=85.5$ for APR4$_{0.7}$.}
\label{fig:BHNS-GW}
\end{figure}

In this article, we investigate a BHNS scenario by comparing the result of one of our BNS simulations with a BHNS model, which contains a 1.39 $M_\odot$ NS described by the APR4 EOS and a non-spinning BH with mass of 1.98 $M_\odot$, i.e., a setup that is 
approximately identical to our $\rm{APR4}_{0.7}$ model. This provides a unique opportunity to compare the properties of the BNS and BHNS mergers. 

Figure~\ref{fig:BHNS-GW} shows the gravitational waveforms observed from the 
rotational axis obtained for the two different models. 
Table~\ref{tab:BHNS-prop} reports the properties of the remnants for the two models. 
The agreement between the GW signals of the two models is remarkable. In 
particular, we should emphasize that both setups have been evolved with different numerical-relativity codes. 
This good agreement between the two waveforms is caused by the fact that in addition to the identical masses of the systems, the tidal deformability is very similar, namely $71.2$ for the BHNS setup and $85.5$ for $\rm{APR4}_{0.7}$.~\footnote{We note that the scenario considered in this work is quite different from the BHNS--BNS comparisons in Ref.~\cite{Hinderer:2018pei}, in which  very low-mass BHNS mergers were considered and larger dephasings have been found.} Moreover, GWs in the merger phase are also virtually indistinguishable, because tidal disruption suppresses the GW emission in both systems. This agreement would make it extremely difficult to differentiate between two scenarios based only on measured tidal contributions, so that only the observed EM signatures could be used to distinguish between the BHNS and BNS origins; see also Refs.~\cite{Hinderer:2018pei,Most:2020exl}.

With respect to the remnant disk and dynamical ejecta mass, we find that the BHNS system produces a disk mass about twice as large as $\rm{APR4}_{0.7}$ (see Tab.~\ref{tab:BHNS-prop}). Because of this larger disk mass, which can generate larger post-merger ejecta, one can expect that the BHNS model will generally produce a brighter kilonova signal, as we will see in the next subsection. We note that the difference of the disk mass between the BNS and BHNS models will be more significant for the equal-mass cases, because the BNS model has a negligible remnant disk mass for $q=1$ while the disk mass is nearly constant for a BHNS model with $q \gtrsim 1/3$~\citep{Hayashi:2020zmn}. 

By contrast the mass of the dynamical ejecta is smaller for our BHNS simulation, i.e., $4.2\times10^{-6}M_\odot$ compared to $5.1\times10^{-4}M_\odot$. Such a large difference is consistent with those found in similar studies \cite{Hinderer:2018pei,Most:2020exl} and is likely a general feature when considering BHNS and BNS mergers with identical component masses and spins. While we note that the masses of the dynamical ejecta derived in the current study are of the order of the numerical uncertainty, we may safely say that the mass of the dynamical ejecta is much smaller than the mass of the disk for both systems.

\begin{table}[t]
\caption{Comparison of the results for BHNS (BHNS$_{0.7}$) and BNS (APR4$_{0.7}$) simulations. The rows refer to the disk mass $M_{\rm disk}$, the dynamical ejecta mass $M_{\rm ej}$, the final BH mass $M_{\rm BH}$ (all given in units of $M_{\odot}$), and the dimensionless spin of the final BH $\chi_{\rm BH}$.}
\label{tab:BHNS-prop}
\centering
\begin{tabular}{l|c|c}
\toprule
 & BHNS$_{0.7}$ & APR4$_{0.7}$
\\ \hline
$M_{\rm disk}$ [$M_{\odot}$] & $8.6\times 10^{-3}$ & $4.8\times 10^{-3}$ \\ 
$M_{\rm ej}$ $[M_{\odot}]$& $4.2\times 10^{-6}$ & $5.2\times 10^{-4}$ \\ 
$M_{\rm BH}$ [$M_{\odot}$] & 3.28 & 3.28 \\ 
$\chi_{\rm BH}$ & 0.75 & 0.74 \\ 
\hline \hline
\end{tabular}
\end{table}

\subsection{Lightcurve Computation}

\begin{figure}[t]
\includegraphics[width=0.49\textwidth]{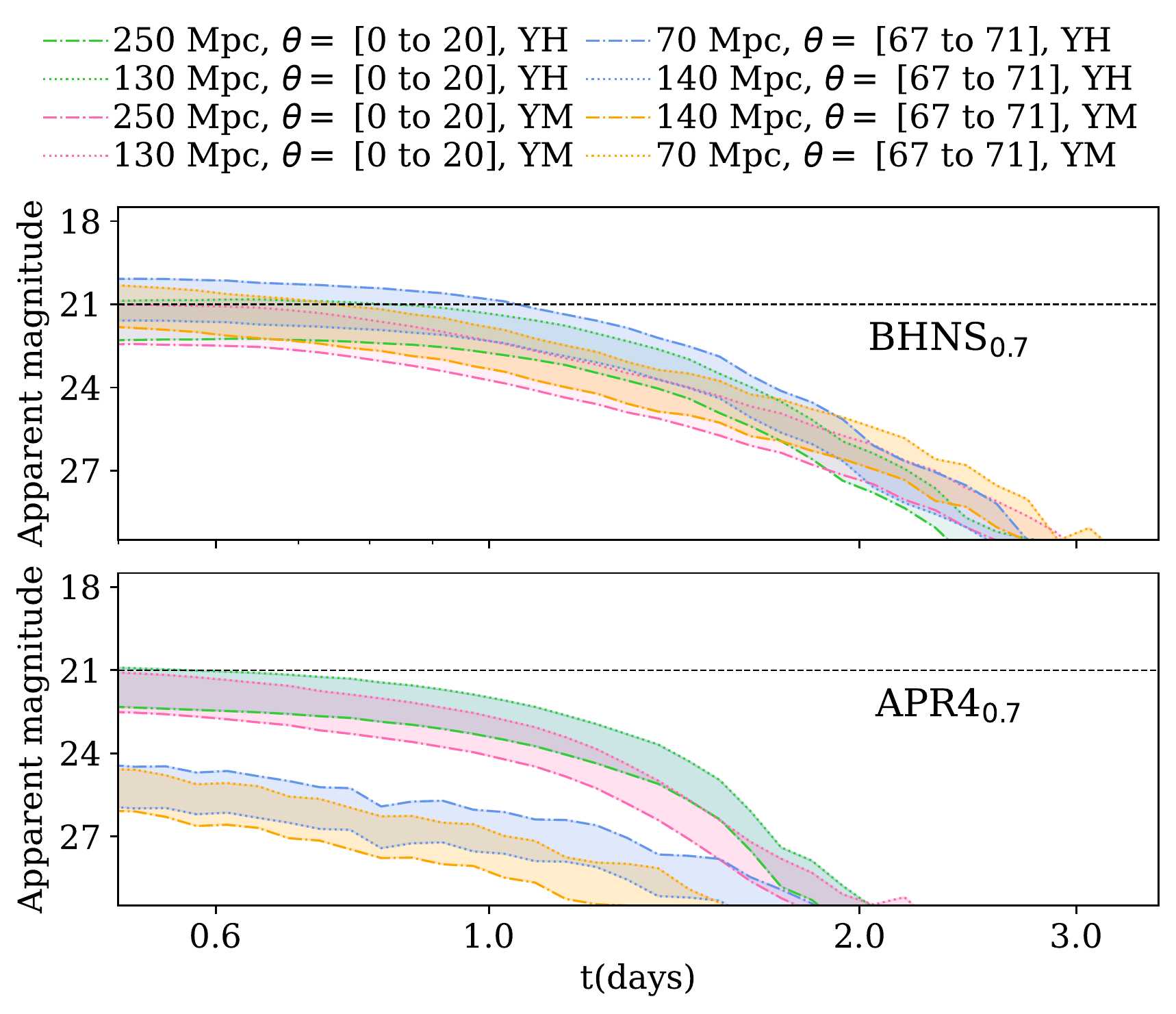}
\caption{Apparent AB magnitude for r-band lightcurve of the kilonovae for the BHNS model (BHNS$_{0.7}$: top panel) and a BNS model (APR4$_{0.7}$: bottom panel).}
\label{fig:BHNS-AB}
\end{figure}

Based on our previous discussion and the methods employed for the BNS models, we compare the lightcurves between the BHNS and $\rm{APR4}_{0.7}$ models in Fig.~\ref{fig:BHNS-AB}. We find that the lightcurves for our BHNS model are much brighter than the ones for the corresponding BNS model, in particular for the case that the kilonova 
is observed from the equatorial direction. In the following, we describe the reason for this. A related discussion can also be found in Ref.~\cite{Kyutoku:2020xka}, in which comparisons of lanthanide-poor and lanthanide-rich outflows are made. 

For BNS cases, the dynamical ejecta have masses comparable with the post-merger ejecta 
in the assumption that 20\% of the remnant disk eventually becomes the post-merger ejecta. Since the dynamical ejecta have non-spherical geometry extended primarily in the equatorial direction, the emission toward the equatorial direction has lower color temperature due to the larger photospheric radius than that observed from the polar direction, and the optical emission (e.g., the r-band lightcurve) is suppressed due to the reddening of the spectrum. Moreover, the emission from the post-merger ejecta located in the center would be blocked by the dynamical ones. This effect also leads to the suppression of the emission towards the equatorial plane (a.k.a. the lanthanide curtain effect~\citep{Kasen:2014toa}). This is basically the reason why fainter emission is observed from the edge-on view than the face-on view in spite of the smaller distance for GW190425. 

On the other hand, the dynamical ejecta mass for the BHNS model is much smaller than 
the disk mass (and hence the hypothetical post-merger ejecta mass). Thus, 
the kilonova emission is dominated by the post-merger ejecta. For such a case, under the assumption of a spherical geometry for the post-merger ejecta, the kilonova emission has very weak viewing angle dependence as suggested by numerical-relativity simulations, e.g., Ref.~\cite{Fujibayashi:2020dvr}. This results in the brighter emission toward the edge-on direction than the face-on direction, simply because the distance ($D$) becomes smaller for the edge-on direction than the face-on direction for GW190425 considered here.
As a result, if the observer is located near the equatorial plane, the kilonova 
for the BHNS model can be much brighter than for the BNS model. 
By contrast, if the observer is located near the rotational axis, the difference in the brightness is not very appreciable because the blocking by the dynamical ejecta is minor. 

\section{Conclusion}

We have presented new results for twelve BNS and one BHNS merger simulations 
with the setup that the chirp mass is similar to that of GW190425 but varing the 
binary mass-ratio and EOSs in order to understand the possible properties of 
GW190425 and to explore whether the non-detection of kilonovae places further 
constraints on the EOS and helps to discriminate between BNS and BHNS merger scenarios. 
We find that we are unable to obtain an EOS-related constraint if the system was an 
equal-mass or 
nearly equal-mass merger, as already pointed out in Ref.~\cite{Kyutoku:2020xka}. However, for unequal masses ($q<0.8$), we find that stiff EOSs 
with the face-on observation of the binary are disfavored 
if we take the missing detection of kilonova signals into account. 
Considering the comparison between BNS and BHNS systems, we find that some 
of the BHNS scenarios of GW190425 would have likely produced a kilonova signal that should have been detected based on the observations of~Ref.~\cite{Coughlin:2019grw}. Specifically, 
BHNS systems that result in negligible dynamical ejecta and in the disk mass 
with $\agt 0.01M_\odot$ should be detected in the GW190425-like event.

Given that GW190425 was basically a one-detector trigger, which led to an imprecise sky localization, had a large distance uncertainty, and had a largely unknown inclination, our constraint should be taken with care. However, for future detections with larger SNRs and observed with more sensitive facilities such as the Vera C. Rubin Observatory, there will be a chance to use kilonova observational results to understand the nature of the compact binary merger and to constrain the source parameters.

\begin{acknowledgments}
  %
  T.D.\ acknowledges financial support through the Max Planck Society. 
This work was in part supported by Grant-in-Aid for Scientific Research (Grant
No. JP18H01213, JP19K14720, JP20H00158) of Japanese MEXT/JSPS and by DFG grant BR 2176/5-1. Numerical computations were performed on Sakura and Cobra clusters at Max Planck Computing and Data Facility, and on FAU's Koko cluster.
  We are also grateful for computational resources provided by the High-Performance Computing Center Stuttgart (HLRS) [project GWanalysis 44189], and the Leibniz Supercomputing Centre (LRZ) [project pn29ba]. W.T. was supported by the National Science Foundation under grants PHY-1707227 and PHY-2011729.
  
  %
\end{acknowledgments}
\bibliography{refs}
\end{document}